\documentclass[twocolumn,npa,showpacs]{revtex4}
\usepackage{amsmath,bm}%
\usepackage{graphicx}%
\usepackage{ulem}

\begin{document}

\draft

\title{Weighted Network of Chinese Nature Science Basic Research
}

\author{Jian-Guo Liu$^1$\footnote{Electronic address: liujg004@yahoo.com.cn}, Zhao-Guo Xuan$^1$, Yan-Zhong Dang$^1$, Qiang Guo$^2$, and Zhong-Tuo
Wang$^1$}
\address{$^1$Institute of System Engineering, Dalian University of Technology,
Dalian 116023, P R China\\ $^2$School of Science, Dalian
Nationalities University, Dalian 116600, P R China}


\begin{abstract} \textnormal{\small {Using the requisition papers of Chinese Nature Science Basic
Research in management and information department, we construct the
weighted network of research areas({\bf WRAN}) represented by the
subject codes. In WRAN, two research areas are considered connected
if they have been filled in at least one requisition paper. The edge
weight is defined as the number of requisition papers which have
filled in the same pairs of codes. The node strength is defined as
the number of requisition papers which have filled in this code,
including the papers which have filled in it only. Here we study a
variety of nonlocal statistics for these networks, such as typical
distances between research areas through the network, and measures
of centrality such as betweenness. These statistics characteristics
can illuminate the global development trend of Chinese scientific
study, it is also helpful to adjust the code system to reflect the
real status more accurately. Finally, we present a plausible model
for the formation and structure of networks with the observed
properties. }}
\end{abstract}


\maketitle
\section{Introduction}
Recently, the topological properties and evolutionary processes of
complex networks are used to describe the relationships and
collective behaviors in many
fields\cite{WS98,BA99,AB02,DM02,New,XFWang01,ADD1}. Some new
analysis methods and topology properties have been proposed by
network analysis. Also it impelled us to study the complex system
from the point of macroscopically view. A network is consisted of a
set of nodes and edges which represent the relationship between any
two nodes. The topological network is denoted by an adjacent matrix
$W=w_{ij}$, if node $i$ connect to node $j$, $w_{ij}=1$; Otherwise,
$w_{ij}=0$. Just because of its simplicity of this description,
network can be used in so many different subjects, such as
collaboration of scientists\cite{Scient1,Scient2,Scient3,Scient4},
Internet networks\cite{Int}, World-Wide Web\cite{BA99}, the
collaborative research and project bipartite network\cite{EU} and so
on. Barber {\it et. al} \cite{EU} studied the collaboration network
consisting of research projects funded by the European Union and the
organizations. They found that the collaboration network has the
main characteristics, such as scale-free degree distribution, small
average distance, high clustering and assortative node correlations.
However, the real systems are far from Boolean structure. The purely
topological characterization will miss important attributes often
encountered in real systems. So to fully characterize the
interactions in real-world networks, weight of links should be taken
into account. In fact, there are already many works on weighted
networks, including empirical studies\cite{WAN1,Li,PP,15,152,BBV2}
and evolutionary models\cite{152,17,18,19,WWX2,WWX,Self}.

The empirical study of weighted network without a naturally given
definition of weight is especially valuable to answer questions such
as how to define a well behavior weight, and to extract structural
information from networks, and what's the role of weight according
to its effects on the structure of the network. We introduce some
metrics that combine in a natural way both the topology of the
connections and the weight assigned to them. These quantities
provide a general characterization of the heterogenous statistical
properties of weights and identify alternative definitions of
centrality, local cohesiveness, and affinity. By appropriate
measurements it is also possible to exploit the correlation between
the weights and the topological structure of the network, unveiling
the complex architecture shown by real weighted networks.

The scientific studies can be considered as being organized within a
network structure, which has a significant influence on the observed
study collective behaviors. The viewpoints of complex networks are
of interest in studying scientific study networks to uncover the
structural characteristics of WRAN. The topological statistics
properties have discussed in Ref.\cite{LDW}. In the fund management
department, such as National Natural Science Foundation of China
(NSFC), the research areas are denoted by the code system, which
have the tree structure to demonstrate the inclusion relation
between the research areas, such as Physics--$>$statistical
physics--$>$complex network. The leave codes of the code system
always represent the research areas more specially. To make the
network reflect the reality more accurately, the nodes are defined
as the codes. Because the scientists can fill in the fund proposal
two codes: the first application code and the second one, then if
one requisition paper filled in two different codes one can consider
that the research work is cross the two research areas. The edge
weight $w_{ij}$ between node $i$ and $j$ is defined as the number of
papers filled in the two codes. The node strength $s_i$ is defined
as the number of requisition papers which have filled code $i$,
including the papers which have filled it only. By this definition,
the network size $N$ is 321 in WRAN from 1999 to 2004. The network
shows all the main characteristics known from other complex network
structure, such as exponential distribution of degree, node weight
and node strength, small average path length, large clustering, and
assortative node correlations. Besides the general interest in
studying the new network, the study could help us to know how the
network structure affects network functions such as knowledge
creation, knowledge diffusion and the collaboration of scientists.
Moreover, the macroscopically analysis can illuminate the global
development trend of Chinese scientific study, it is also helpful to
adjust the code system to reflect the real status more accurately.

\section{Measurement of weight and basic statistical results}
\begin{figure}
  \begin{center}
       \center \includegraphics[width=8cm]{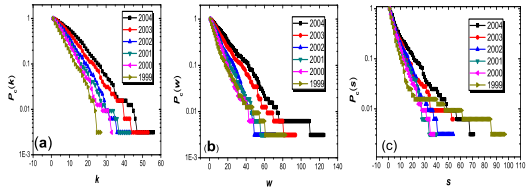}
       \caption{(Color online) Characteristics of WRAN, such as the distributions of
degree, node weight and node strength.}\label{WFig1}
 \end{center}
\end{figure}

Now we turn to the effects of weight on the structure of weighted
networks. First, the interaction weight $w_{ij}$ is define as the
number of requisition papers which have filled in code $i$ and code
$j$. The strength $s_i$ of node $i$ is defined as
\begin{equation}
s_i=\sum_{j\in \Gamma_i}w_{ij}+\eta_i,
\end{equation}
where $\Gamma_i$ is the neighbor node set of node $i$ and the
fitness $\eta_i$ is the number of requisition papers which filled in
the code $i$ only. The weight $w_i$ of node $i$ is defined as
\begin{equation}
w_i=\sum_{j\in \Gamma_i}w_{ij}.
\end{equation}
This quantity measures the strength of nodes in terms of the total
weight of their connections. The distributions of degree, node
weight and node strength are demonstrated in Fig.\ref{WFig1}. The
probability distribution $P(s)$ that a node has strength $s$ is
exponential distribution, and the functional behavior exhibits
similarities with the degree distribution $P(k)$ (see
Fig.\ref{WFig1}). The largest strength nodes have been listed in
Table 1.

\begin{figure}
\begin{center}
\center \includegraphics[width=8cm]{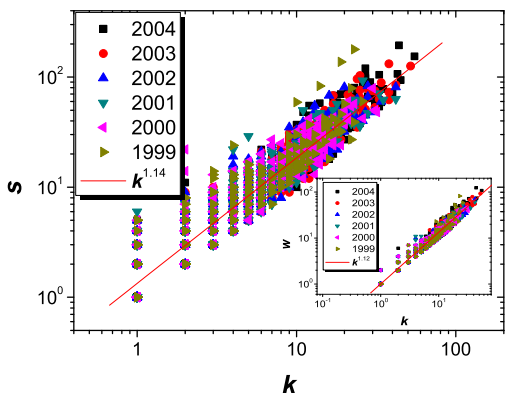} \caption{(Color
online) Average strength $s(k)$ as function of the degree $k$ of
nodes from 1999 to 2004. The inset figure shows the relationship
between the average node weight $w(k)$ and the degree
$k$.}\label{WFig2}
\end{center}
\end{figure}

A precise functional description of the exponential distributions
may be very important for understanding the network evolution and
will be deferred to future analysis. To shed more light on the
relationship between the node strength and degree, we investigate
the dependence of $s_i$ on $k_i$. We find that the average strength
$s(k)$ and weight $w(k)$ of nodes with degree $k$ increase with the
degree as
\begin{equation}
s(k)\sim k^{\beta_{sk}},\ \ w(k)\sim k^{\beta_{wk}}.
\end{equation}
The real data follows the power-law behavior with exponent
$\beta_{sk}=1.14\pm 0.02$ and $\beta_{wk}=1.12\pm 0.01$(see
Fig.\ref{WFig2}). The two exponents denote anomalous correlations
between the number of paper which has filled in one node and the
number of its connections, and imply that the strength and weight of
nodes grows faster than their degree and the weight of edges
belonging to highly connected nodes tends to have a higher value.
This tendency denotes a strong correlation between the strength,
node weight and the topological properties in WRAN. The difference
between $\beta_{sk}$ and $\beta_{wk}$ implies that the larger degree
a node is, the more fitness $\eta_i$ it has.

\begin{widetext}
\begin{center}
{Table 1, The hub nodes of WRAN and their strength from 1999 to
 2004.}
  \begin{tabular}{c c c}\hline
   Year      &  Hub nodes                              & $s$ \\ \hline
   1999      & Corporation theory                      & 178 \\ 
   2000      & Macroscopical economy management and stratagem & 79 \\ 
   2001      & Corporation stratagem management        & 93  \\
   2002      & Computer network, distributed computer system(CNDCS) & 83 \\
   2003      & CNDCS                                   & 132 \\
   2004      & CNDCS                                   & 194\\ \hline
 \end{tabular}
\end{center}
\end{widetext}

\subsection{Distance and Centrality}
\begin{figure}
\begin{center}
\center \includegraphics[width=8cm]{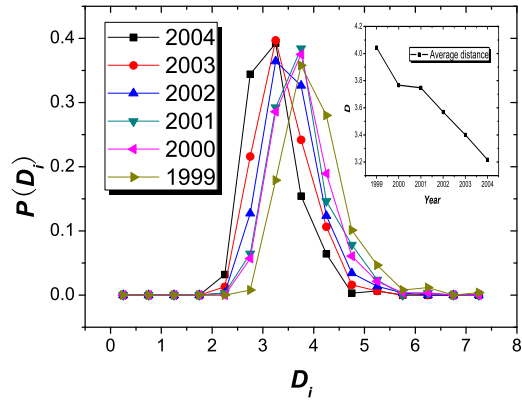} \caption{(Color
online) The topological $D_i$ distributions from 1999 to 2004 obey
Passion distribution. The inset figure shows the average distance
$D$ from 1999 to 2004.}\label{WFig3}
\end{center}
\end{figure}

\begin{figure}
\begin{center}
\center \includegraphics[width=8cm]{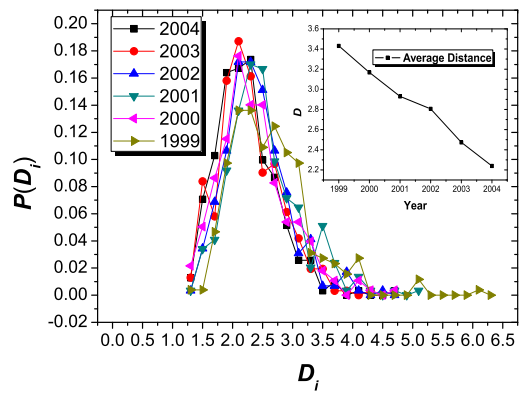} \caption{(Color
online) The weighted $D_i$ distributions of WRAN from 1999 to 2004
obey Passion distribution. The inset figure shows the average
distance $D$ of weighted RAN from 1999 to 2004.}\label{WFig4}
\end{center}
\end{figure}

\begin{figure}
\begin{center}
\center \includegraphics[width=8cm]{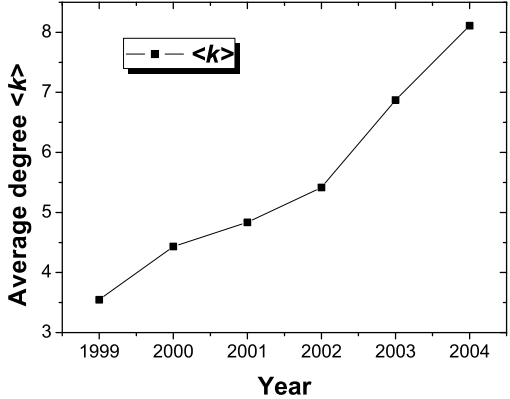} \caption{(Color
online) Average degree $\langle k\rangle$ from 1999 to 2004, which
is increase almost 2 times from 1999 to 2004.}\label{WFig5}
\end{center}
\end{figure}

Shortest path play an important role in the transport and
communication within a network, it have also played an important
role in the characterization of the internal structure of a
network\cite{Dis1,Dis2}. The average distance, denoted by
$D=\frac{1}{N(N-1)}\sum_{ij}d_{ij}$, represent all the average
shortest path lengths of a network in which  the entry $d_{ij}$ is
the shortest path length from node $i$ to node $j$. It should be
noticed that all the network nodes are not all connected in the six
years. The largest connected group has 256, 279, 293, 290, 309 and
310 nodes, respectively. The average distance is discussed on the
largest connected group. The ability of two nodes, $i$ and $j$, to
communicate with each other depends on the length of the shortest
path $d_{ij}$ between them. The average distance from node $i$ to
all other nodes is defined as
\begin{equation}
D_i=\frac{1}{N-1}\sum_{j=1,j\neq i}^Nd_{ij}.
\end{equation}
In the Boolean structure network, if nodes $i$ and $j$ are
connected, $d_{ij}=1$. In WRAN, the larger edge weight $w_{ij}$ is,
the closer relationship between the two nodes have. Thus, the
weighted distance $d_{ij}$ is taken $d_{ij}=1/w_{ij}$. The weighted
shortest path length $d_{ij}$ of WRAN is defined as the smallest sum
of the distance throughout all the possible paths in the network
from node $i$ to $j$. Figure \ref{WFig3}, \ref{WFig4} demonstrate
the topological and weighted $D_i$ distributions from 1999 to 2004
respectively, which both obey Passion distribution. From the two
figures, we can obtain that most nodes' average distance $D_i$ are
around 3.5 and 2.2 in topological and weighted network,
respectively. The nodes belonging to the left part of Passion
distribution are very important to the network, because their
average distance to all other nodes is very small. The two inset
figures show that the average distance $D$ of topological and
weighted network decreases with time. This may caused by the
increase of the average degree $\langle k\rangle$(See Fig.
\ref{WFig5}). Since the number of requisition papers $E$ can be
obtained from the equation $E=N\langle k\rangle$ approximately, the
real reason why the average distance decrease may lie in the
increasing number of requisition papers.

\subsection{Average Clustering coefficient}
\begin{figure}
\begin{center}
\center \includegraphics[width=8cm]{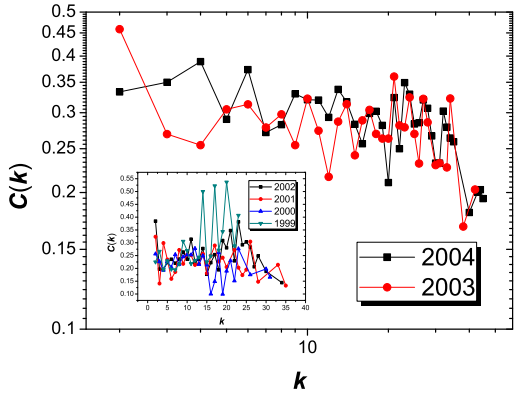} \caption{(Color
online)The topological clustering coefficient vs time from 1999 to
2004.}\label{WFig6}
\end{center}
\end{figure}

The {\it local clustering coefficient} of node $i$, denoted by
$C_i$, is a measure of the connectedness between the neighbors of
the node, which is called {\it transitivity} in the social
network\cite{WS98,Dis1}. If a node $i$ has a link to node $j$ and
node $j$ has a link to node $k$, then a measure of transitivity in
the network is the probability that node $i$ has a link to node $k$.
Let $k_i$ denote the degree of node $i$, and let $E_i$ denote the
number of link between the $k_i$ neighbors. Then, for an undirected
network, the quantity\cite{WS98}
\begin{equation}
C_i=\frac{2E_i}{k_i(k_i-1)}
\end{equation}
is the ratio of the number of links between a node's neighbors to
the number of links that can exist. The clustering coefficient $C$
is defined as $C=1/N\sum_{i=1}^NC_i$. In WRAN, the clustering
coefficient indicates the probability that a node connects to its
$2$nd nearest neighbors. Figure \ref{WFig6} presents the statistic
result of $C(k)\sim k$. From Fig.\ref{WFig6}, we can obtain that
there are no correlation between $C(k)$ and $k$ before 2003, but the
correlation emerged since 2003, which is a characteristic of
hierarchical network. The reason may lie in the fact that the code
system has been adjusted around 2002. This result indicates that the
rectification make the relationship of the subject codes becoming
more clear.

\begin{figure}
\begin{center}
\center \includegraphics[width=8cm]{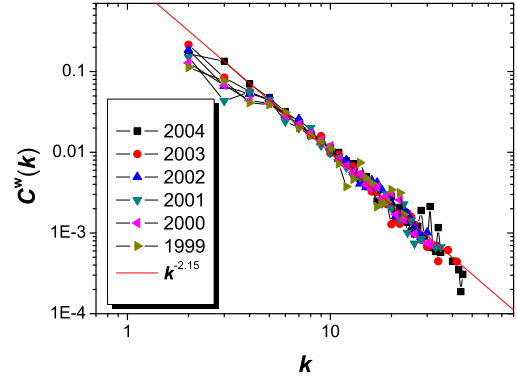} \caption{(Color
online)The weighted clustering coefficient vs time from 1999 to
2004.}\label{WFig7}
\end{center}
\end{figure}

\begin{figure}
\begin{center}
\center \includegraphics[width=8cm]{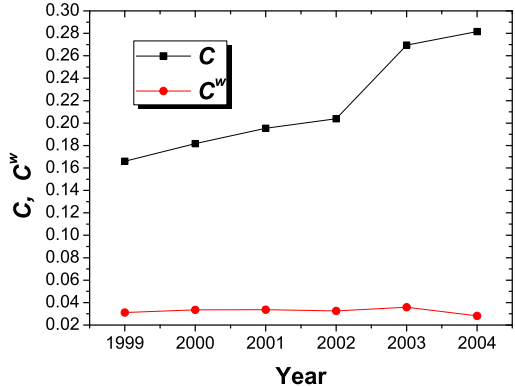} \caption{(Color
online) Topological and weighted clustering coefficient of WRAN from
1999 to 2004.}\label{WFig8}
\end{center}
\end{figure}

The weighted clustering coefficient is defined as
\begin{equation}\label{WF1}
c_i^w=\frac{1}{s_i(k_i-1)}\sum_{j,h\in
\Gamma_i}\frac{w_{ij}+w_{ih}}{2}a_{ij}a_{ih}a_{jh}.
\end{equation}
This coefficient is a measure of the local cohesiveness that takes
into account the importance of the clustered structure on the basis
of the amount of traffic or interaction intensity actually found on
the local triplets. Indeed, $c_i^w$ counts for each triplet formed
in the neighborhood of the node $i$ the weight of the two
participating edges of the node $i$. In this way we are considering
not just the number of closed triplets in the neighborhood of a node
but also their total relative weight with respect to the strength of
the node.
Consistently, the $c_i^w$ definition recovers the topological
clustering coefficient in the case that $w_{ij}$ is constant and
$\eta_i=0$. Next we define $C^w$ and $C^w(k)$ as the weighted
clustering coefficient averaged over all nodes of the network and
over all nodes with degree $k$, respectively. These quantities
provide global information on the correlation between weights and
topology, especially by comparing them with their topological
analogs. Figure \ref{WFig7} presents the power-law correlations
$C^w(k)\sim k^{\alpha}$ between $C^w(k)$ and degree $k$, where
$\alpha=-2.15\pm 0.06$, which may be caused by the introduction of
node fitness $\eta_i$. Because the larger the degree $k$ is the
larger $\eta_i$ would have, the denominator of Equ. (\ref{WF1})
would become more larger, then $C^w(k)$ would become small.
If replace $s_i$ of Equ.(\ref{WF1}) with $k_i$, we get the
definition of weighted clustering coefficient presented in
Ref.\cite{15}. Figure \ref{WFig8} presents the relationship between
$C^w$ and $C$ of WRAN. The fact $C^w<C$ signals a network in which
the topological clustering is generated by edges with low weight or
by nodes with larger fitness. In this case the clustering has a
minor effect in the organization of the network because the largest
part of the interactions is occurring on edges not belonging to
interconnected triplets. The figure also indicates that $C$ increase
with time, while $C^w$ keep constant. Interestingly, $C$ increase
dramatically about 10 percent from 2002 to 2003. This change is
consistent with the correlation $C(k)\sim k$.

\begin{figure}
\begin{center}
\center \includegraphics[width=8cm]{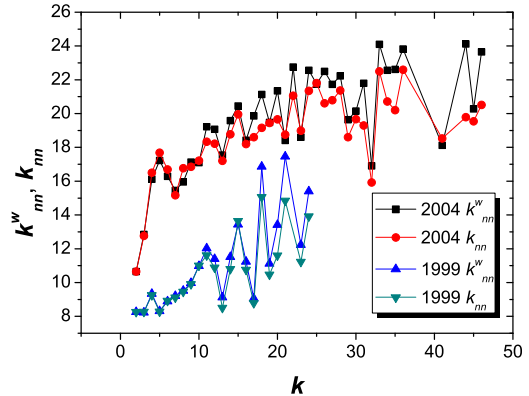} \caption{(Color
online)Topological and weighted average nearest-neighbors degree of
WRAN of 1999 and 2004.}\label{WFig9}
\end{center}
\end{figure}

Along with the weighted clustering coefficient, we introduce the
weighted average nearest-neighbors degree\cite{15}, defined as
\begin{equation}
k_{nn,i}^w=\frac{1}{s_i}\sum_{j=1}^Na_{ij}w_{ij}k_j.
\end{equation}
In this case, we perform a local weighted average of the
nearest-neighbor degree according to the normalized weight of the
connecting edges, $w_{ij}/s_i$. This definition implies that if the
edges with the larger weight are pointing to the neighbors with
larger degree, $k_{nn,i}^w>k_{nn,i}$; In the opposite case
$k_{nn,i}^w<k_{nn,i}$. Thus, $k_{nn,i}^w$ measures the effective
affinity to connect with high- or low-degree neighbors according to
the magnitude of the actual interactions. Moreover, $k_{nn}^w (k)$
marks the weighted assortative or disassortative properties
considering the actual interactions among the system¡¯s elements.
Figure \ref{WFig9} presents the topological and weighted average
nearest-neighbors degree of 1999 and 2004, which demonstrate that
$k_{nn,i}^w>k_{nn,i}$ and both of them have the trend of increasing
with the degree $k$.

\begin{figure}
\begin{center}
\center \includegraphics[width=8cm]{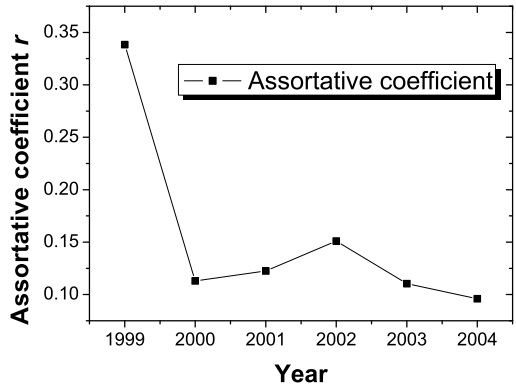} \caption{(Color
online)Assortative coefficient vs time of WRAN.}\label{WFig10}
\end{center}
\end{figure}

The positive assortative coefficient $r$, which is presented by Ref.
\cite{N2002,NP2003}, of WRAN has presented in Fig.\ref{WFig10},
which means that the nodes with higher degree would like to connect
each other. Figure \ref{WFig2} told us that the nodes, whose degree
is large, must have larger strength. Then, the nodes with more
strength would like to connect each other.

\subsection{Betweenness}
\begin{figure}
\begin{center}
\center \includegraphics[width=8cm]{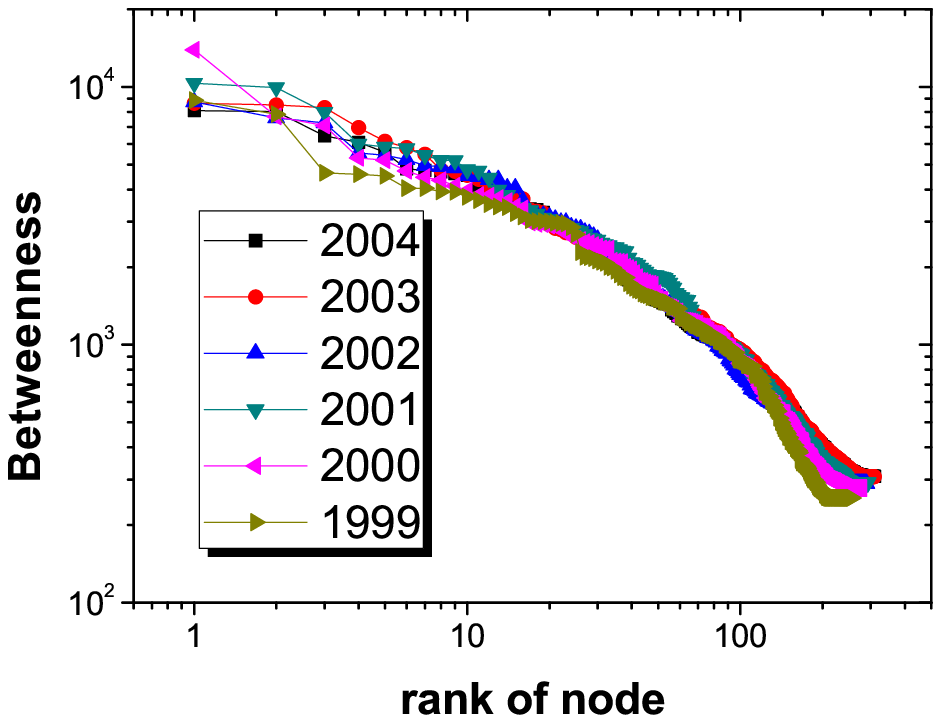} \caption{(Color
online)Zipf plots of node betweenness for topological WRAN from 1999
to 2004.}\label{WFig11}
\end{center}
\end{figure}
\begin{figure}
\begin{center}
\center \includegraphics[width=8cm]{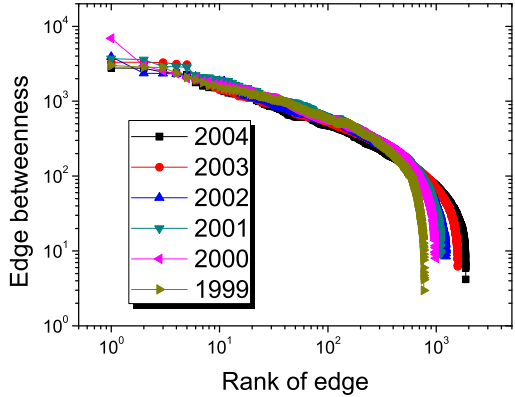} \caption{(Color
online)Zipf plots of edge betweenness for topological WRAN from 1999
to 2004.}\label{WFig12}
\end{center}
\end{figure}
The communication of two non-adjacent nodes, called $j$ and $k$,
depends on the nodes belonging to the paths connecting $j$ and $k$.
Consequently, the definition {\it node betweenness} is present to
measure the relevance of a given node by counting the number of
geodesics going through it. The betweenness is one of the standard
measures of node centrality. The betweenness $b_i$ of node $i$, is
defined as\cite{Dis1,Dis2,Betw1,Betw2}
\begin{equation}
b_i=\sum_{j,k=1,j\neq k}^N\frac{n_{jk}(i)}{n_{jk}},
\end{equation}
where $n_{jk}$ is the number of shortest paths connecting $j$ and
$k$, while $n_{jk}(i)$ is the number of shortest paths connecting
$j$ and $k$ and passing through $i$. This quantity is an indicator
of which node is the most influential one in the network is. The
nodes with highest betweenness also result in the largest increase
in typical distance between others when they are removed. The nodes
with largest betweenness have listed in Table 2. These nodes are the
most important one for information transitivity.

The {\it edge betweenness} is defined as the number of shortest
paths between pairs of nodes that run through that edge
\cite{Betw3}.

\begin{center}
{Table 2 The node wich has largest betweenness from 1999 to
 2004.
  \begin{tabular}{c c}\hline
             &  The node with largest betweenness  \\\hline 
   1999      & Computer-aided design               \\
   2000      & Intelligent information processing  \\
   2001      & Management information system       \\
   2002      & Management information system       \\
   2003      & Artificial intellegence(AI)         \\
   2004      & Intelligent information processing(IIP) \\\hline
 \end{tabular}
 }
\end{center}

\section{A Mutual Selection Model}
\begin{figure}
\begin{center}
\center \includegraphics[width=8cm]{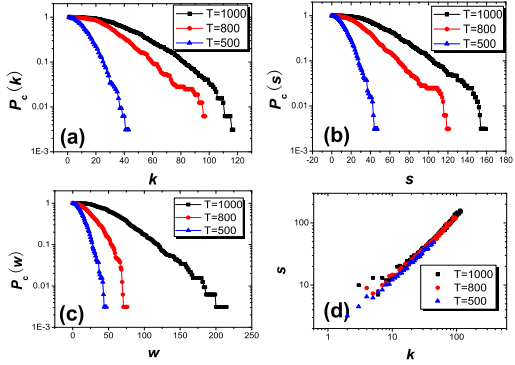} \caption{(Color
online) Simulated distributions of degree, node strength and edge
weight to different time step $T$. (d) give the relationship between
$k$ and $s$. }\label{WFig31}
\end{center}
\end{figure}

\begin{figure}
\begin{center}
\center \includegraphics[width=8cm]{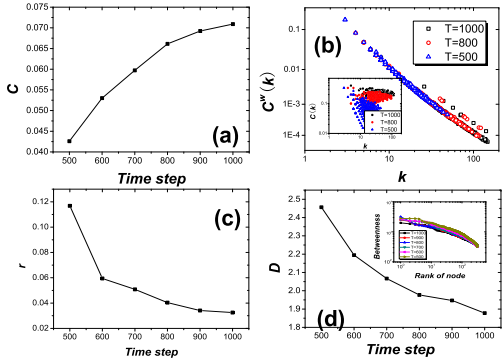} \caption{(Color
online) Simulated numerical results of $C$, $D$ and $r$ to different
$T$, and the relationship of $C(k)\sim k$ and $C^w(k)\sim k$, which
.}\label{WFig32}
\end{center}
\end{figure}

In this section, we present a mutual selection model (MSM) to
compare with WRAN. Inspired by the fitness $\eta_i$ and the mutual
selection mechanism, the model is defined as following. The model
starts from $N$ isolated nodes, each with an initial attractiveness
$s_0$. In this paper, $s_0$ is set to be 1. At each time step, every
node strength of the network would increase by 1 with the
probability $p$; With the probability $(1-p)$, each existing node
$i$ selects $m$ other existing nodes for potential interaction
according to the probability Equ. (\ref{F5.1}). Here, the parameter
$m$ is the number of candidate nodes for creating or strengthening
connections, $p$ is the probability that a node would enhance
$\eta_i$ by 1.
\begin{equation}\label{F5.1}
\Pi_{i\rightarrow j}=\frac{s_j}{\sum_{k(k\neq i)} s_k}.
\end{equation}
where $s_i=\sum_{j\in \Gamma(i)}w_{ij}+\eta_i$. If a pair of
unlinked nodes is mutually selected, then an new connection will be
built between them. If two connected nodes select each other, then
their existing connection will be strengthened, i.e., their edge
weight will be increased by 1. We will see that the model can
generate the observed properties of WRAN. When $p=0.01$ and $m=5$,
the numerical results to different time step $T$ are demonstrated in
Fig. \ref{WFig31}-\ref{WFig33}. Figure \ref{WFig31}. (a)-(c) give
the exponential distributions of degree, node strength and edge
weight. Figure \ref{WFig31}. (d) demonstrate the power-law
relationship between degree $k$ and node strength $s$. Figure
\ref{WFig32} demonstrates the increasing trend of $C$, decreasing
trend of $D$ and $r$ and the $C^w(k)\sim k^\lambda$ relationship.
From the inset of Fig.\ref{WFig32}. (b), one can see that when the
time step $T$ is very small, there is no correlation between $C(k)$
and $k$, while when $T$ is become large, the correlation emerge,
which consistent with $C(k)\sim k$ of WRAN. Figure \ref{WFig32}. (b)
gives the power-law relationship $C^w(k)\sim k^{\alpha}$, where
$\alpha=1.11\pm 0.05$, which also consistent with the one of WRAN.
The inset of Figure \ref{WFig32}.(d) gives the Zipf plots of node
betweenness to different time step $T$. All of the above structural
characters of MSM are consistent with the ones of WRAN
approximately, which indicate that the mutual selection mechanism
and the probability $p$ may be the evolving mechanism of WRAN.

\section{Conclusions and Discussions}
We have studied the Chinese Nature Science Basic Research in
management and information department from weighted network point of
view. To describe the status of WRAN more accurately, the
requisition papers which have filled in only one subject code is
also considered, which is defined as node fitness. We have looked at
a variety of nonlocal properties of our networks.

Using this measure we have added weighting to WRAN and used the
resulting networks to find which code have the largest strength, the
shortest average distance to others. Generalization of the
clustering coefficient and betweenness calculations to these
weighted networks is also straightforward. The statistic
characterization give the following conclusions
\begin{description}
\item[(1).] The code system have adjusted around 2002 and the
correlation between $C(k)\sim k$ emerges since 2003.

\item[(2).] The topological and weighted distance decrease with
time, while the clustering coefficient increases with
time.

\item[(3).] The distributions of degree, edge weight and node strength
have exponential form.

\item[(4).] The larger the node degree is, the larger fitness it
would be.

\item[(5).] WRAN is assortative, which means that the node with large
strength would like to connect each other.
\end{description}

In terms of structural characteristics of WRAN, the present analysis
yields a plausible model. Based on the mutual selection mechanism
and the probability $p$ that one node would increase its strength
without creating new connectivity with others, we presented MSM
model. Most of the structural characters of MSN are consistent with
the ones of WRAN.

The calculations presented in this paper inevitably represent only a
small part of the investigations that could be conducted using large
network data sets such as these. We hope, given the high current
level of interest in network phenomena, that others will find many
further uses for these data.

\section*{Acknowledgements}
The authors thank W. -X. Wang and T. Zhou for their valuable
comments and suggestions. This work has been partly supported by the
Natural Science Foundation of China under Grant Nos. 70431001 and
70271046.

\end{document}